\documentclass[twocolumn,english,pra,superscriptaddress]{revtex4-1}
\usepackage[utf8]{inputenc}
\usepackage{xcolor}
\usepackage{babel}
\usepackage{bm}
\usepackage{graphicx}
\usepackage{amsmath}
\usepackage{amssymb}
\usepackage{empheq}
\usepackage{times}
\usepackage{physics}
\usepackage{color}

\usepackage[colorlinks=true,citecolor=blue]{hyperref}
\begin{document}

\title{Speed limits of two-qubit gates with qudits}

\author{Bora Basyildiz}
\email{bbasyildiz@mines.edu}
\affiliation{Quantum Engineering Program, Colorado School of Mines, Golden, Colorado 80401, USA}

\author{Casey Jameson}
\affiliation{Department of Physics, Colorado School of Mines, Golden, Colorado 80401, USA}



\author{Zhexuan Gong}
\email{gong@mines.edu}
\affiliation{Quantum Engineering Program, Colorado School of Mines, Golden, Colorado 80401, USA}
\affiliation{Department of Physics, Colorado School of Mines, Golden, Colorado 80401, USA}
\affiliation{National Institute of Standards and Technology, Boulder, Colorado 80305, USA}

\begin{abstract}
The speed of elementary quantum gates ultimately sets the limit on the speed at which quantum circuits can operate. For a fixed physical interaction strength between two qubits, the speed of any two-qubit gate is limited even with arbitrarily fast single-qubit gates. In this work, we explore the possibilities of speeding up two-qubit gates beyond such a limit by expanding our computational space outside the qubit subspace, which is experimentally relevant for qubits encoded in multi-level atoms or anharmonic oscillators. We identify an optimal theoretical bound for the speed limit of a two-qubit gate achieved using two qudits with a bounded interaction strength and arbitrarily fast single-qudit gates. In addition, we find an experimentally feasible protocol using two parametrically coupled superconducting transmons that achieves this theoretical speed limit in a non-trivial way. We also consider practical scenarios with limited single-qudit drive strengths and off-resonant transitions. For such scenarios, we develop an open-source, machine learning assisted, quantum optimal control algorithm that can achieve a speedup close to the theoretical limit with near-perfect gate fidelity. This work opens up a new avenue to speed up two-qubit gates when the physical interaction strength between qubits cannot be easily increased while extra states outside the qubit subspace can be well controlled.
\end{abstract}

\maketitle
\raggedbottom

\section{Introduction}\vspace{-5pt}
As noisy intermediate-scale quantum computers \cite{Preskill_NISQ,IBM_NISQ} have finite coherence times and imperfect gate fidelities, designing quantum gates with faster speed and higher fidelity has been a major focus in quantum computing  \cite{MIT_Fidelity,Noh2023_Param,Srinivas2021IonGates}. Tremendous effort from both academia and industry has been made on improving gate fidelities. This improvement, when coupled with the recent development of low-cost error correction codes \cite{Breuckmann2021,bravyi2023,xu2023}, means that we may no longer be far from fault-tolerant quantum computers. Therefore, increasing the speed of a quantum computer would become a more important challenge in the long term. The most direct way to address this challenge is to achieve faster quantum gates, especially two-qubit gates since they are usually the speed bottleneck for most hardware platforms \cite{SC_Coupl,NeutralAtom2,FluxmonsCoher}. However, the speed of two-qubit gates is fundamentally limited by the interaction strength between the physical qubits \cite{SpeedLim}. An upper limit of such interaction strength usually exists for a given hardware design, posing a strong constraint on the speed of quantum computers.

One possible way to alleviate this constraint is to speed up two-qubit gates using extra states present in physical systems that carry the qubits. For example, if qubits are encoded in atomic states, there often exist low-energy states other than the qubit states that can be well controlled and have long coherence times \cite{Low2020qudits,low2023control,Hrmo2023quditcontrol}. Alternatively, for qubits encoded in anharmonic oscillators such as superconducting qubits, one can often control and measure higher energy states above the qubit states \cite{long2021universal,Fisher2023quditransmons,cao2023emulating,FluxmonsCoher}. As a result, we can expand the computational space from qubits to qudits (where $d$ stands for the dimension of the local Hilbert space). This allows us to go beyond the usual speed limit for two-qubit gates \cite{SpeedLim} without increasing the physical interaction strength. While we can also perform universal quantum computing using qudits as the native information carriers \cite{Low2020qudits,Wang_2020,Luo2014qudits,Chi2022qudits,yeh2023scaling,Roy2023scaling}, this approach would require significant effort in developing compatible quantum algorithms \cite{Shors_Algo,Grovers_Algo,HHL}, efficient quantum circuit compilations \cite{Wille2020circcomp,Shan2023circcomp,quetschlich2023compiler}, and quantum error correction codes \cite{Steane_ErrCorrec,Calderbank1996errorcorrec,Roffe_2019}. Here we only intend to use qudits effectively as qubits for quantum computing, but with a potential speed advantage over the use of strictly qubits. More explicitly, we require the state of the system to be within the qubit subspace before and after each quantum gate, such that states beyond the qubit subspace can only be populated in the middle of the gate implementation. In fact, many existing quantum gate designs already involve states outside the qubit subspace in a similar way \cite{Qutrit1,YaleSC2,DRAG}. A simple example is a single-qubit gate achieved by a Raman transition. Gates designed this way do not require end users to consider states outside the qubit subspace, which greatly simplifies the design of quantum circuits.

The idea of using extra states beyond the qubit subspace to speed up two-qubit gates has been recently explored in a system made of two coupled weakly anharmonic superconducting qubits \cite{Ashhab}. Here we ask a more general question: What is the fundamental speed limit of a two-qubit gate using two qudits with a bounded interaction strength? For $d=2$, the answer to this question is known for any given two-qubit gate and the form of interaction, provided that arbitrarily fast single-qubit gates are available \cite{SpeedLim}. For $d>2$, no similar answer exists to our best knowledge. Ref.\,\cite{Ashhab} studies such a speed limit for a specific two-qudit Hamiltonian. But since the study relies on numerical optimization, no well-established speed limit was found. In this work, we provide more in-depth answers to the above question by first deriving a rigorous theoretical bound on the speed of certain two-qubit gates using a general two-qudit Hamiltonian. We show with explicit protocols that this bound cannot be generically improved and is thus optimal. 

Next, we argue that for physical Hamiltonians, the speedup of two-qubit gates due to extra states in qudits comes from two distinct sources: (1) larger coupling strengths between higher energy states of each qudit and (2) constructive interference among multiple states beyond the qubit subspace. For example, the speedup observed in Ref.\,\cite{Ashhab} is largely due to (1), and we provide an $O(d)$ upper bound of such speedup. Here we construct explicit protocols that can achieve speedup from either or both of the two sources. Importantly, we show that for $d=3$, these protocols can be readily implemented using two parametrically coupled superconducting transmon qubits \cite{RayParam}. The use of just one additional excited state for each transmon qubit can already lead to a two-qubit iSWAP gate 3 times its maximum speed with only two qubits. We further show that this gate protocol is time optimal as no faster iSWAP gate can be achieved based on the same interaction Hamiltonian.

In practice, it is hard to achieve the theoretical speed limit mentioned above as it requires arbitrarily fast single-qudit gates and no errors due to off-resonant drives or couplings. To address this concern, we develop an open-source quantum optimal control algorithm combining the GRAPE algorithm \cite{GRAPE} with techniques in machine learning to optimize the fidelity of a target two-qubit gate with two coupled qudits. We optimize the pulse shapes of the drives on both qudits and identify the shortest gate time for a gate fidelity above a desired threshold. As an example, we focus on designing a speed-optimized iSWAP gate using two coupled qutrits as discussed above. Our optimal control algorithm generates gates with \textgreater $99.99\%$ fidelity at a speed close to the theoretical speed limit, assuming the maximum Rabi frequency of the drives is about an order of magnitude larger than the interaction strength. If off-resonant transitions are taken into account, the gate fidelity will be lower (around $99\%$) for the gate close to the theoretical speed limit, but we can slow down the gate by a small amount to achieve \textgreater $99.9\%$ gate fidelity.

It is worth mentioning that the open-source repository we developed for our optimal control algorithm can be useful for a wide range of problems in quantum gate designs. Apart from its high efficiency due to the use of state-of-art machine learning techniques, it can handle a general time-dependent two-qudit Hamiltonian and multiple drives at different frequencies, including the off-resonant transitions they may induce. Our repository thus goes beyond several existing counterparts for GRAPE-based algorithms \cite{qutip,Schuster1,Schuster2,LLNL1,LLNL2}.  

The paper is organized as follows. In Section II, we study the theoretical speed limit of two-qubit gates achieved with two coupled qudits. We discuss explicit protocols that can saturate the theoretical speed limit and explain the two different sources of speedup for physical Hamiltonians. In Section III, we introduce our quantum optimal control algorithm and apply it to find speed-optimized two-qubit gates using two parametrically coupled transmon qutrits. Section IV concludes the paper with an outlook.

\section{Theoretical Speed limit}

We first derive a theoretical speed limit for a given two-qubit gate assuming a bounded interaction strength between two qudits and arbitrarily fast single-qudit gates. The Hamiltonian we consider is written in the following general form:
\begin{equation} \label{Ht}
    H(t) = H_1(t) + H_2(t) + H_c(t)
\end{equation}
where $H_1(t)$ and $H_2(t)$ are the Hamiltonians that act only on the qudit 1 and qudit 2 respectively. $H_c(t)$ denotes the interaction between the two qudits. We upper bound the interaction strength via $\lVert H_c(t) \rVert \le J$, where $\lVert \cdot \rVert$ denotes the operator norm (largest singular value of the operator) and $J$ is a constant that only depends on the dimension $d$ of each qudit. We do not limit $\lVert H_{1,2}(t) \rVert$, which allows for arbitrarily fast single-qudit gates.

To find the speed limit for the above Hamiltonian $H(t)$ to generate a given two-qubit gate, we first recall results in the study of quantum speed limits \cite{QST} that bound the minimum time $t_{\perp}$ to evolve an arbitrary quantum state to an orthogonal state. For example, the Mandelstam-Tamm (MT) inequality \cite{MT} shows that for any time-independent Hamiltonian $H$, 
\begin{equation}
    t_\perp \geq \frac{\pi}{2\Delta E} \label{MTEq}
\end{equation}
where $\Delta E = \sqrt{\langle H^2 \rangle - \langle H \rangle^2}$ is the energy uncertainty of the initial state. We assume $\hbar=1$ throughout this paper.

The MT inequality can be generalized to accommodate a time-dependent Hamiltonian $H(t)$ and an arbitrary overlap between an initial state $|\psi(0)\rangle$ and a final state $|\psi(T)\rangle$ \cite{MT}, which reads:
\begin{equation} \label{MTtime}
   \int_0^{T} \Delta E(t) dt \le \arccos{|\langle \psi(0)|\psi(T)\rangle|}
\end{equation}
where $\Delta E(t)$ is the uncertainty of $H(t)$ in the state $|\psi(t)\rangle$. 

We cannot directly apply the MT inequality or its above generalization to bound the speed of a two-qubit gate. This is because we have allowed for arbitrarily fast single-qudit gates. Thus, it takes zero time to evolve from any state to an orthogonal state. To address this issue, we need to separate the single-qudit Hamiltonians $H_{1,2}(t)$ from the interaction Hamiltonian $H_c(t)$ in Eq.\,\eqref{Ht}. This can be done by introducing the following interaction picture Hamiltonian for $H_c(t)$:
\begin{equation} \label{HcI}
        H_I(t) = U^{\dagger}_{H_1}(t) U^{\dagger}_{H_2}(t) H_c(t) U_{H_1}(t) U_{H_2}(t)
\end{equation}
where we define the time evolution operator $U_G(t) \equiv \mathcal{T} e^ {-i \int_0^t  G(t^{\prime}) dt^{\prime}}$ for a general time-dependent Hamiltonian $G(t)$. Since $U_G(t)$ is a unitary operator, it's easy to see $\lVert H_I(t) \rVert = \lVert H_c(t)\rVert$.

Our goal is to implement a given two-qubit target gate $\mathcal{U}$ in time $T$ in the two-qubit subspace of the two qudits. This means we require $U_H(T) = \begin{pmatrix}\mathcal{U} & 0\\ 0 & \tilde{\mathcal{U}} \end{pmatrix}$, where $\tilde{\mathcal{U}}$ is an arbitrary unitary outside the two-qubit subspace. Such block-diagonal $U_H(T)$ ensures that as long as the initial state of the system belongs to the two-qubit subspace, the final state does so too.

We can use the generalized MT inequality in Eq.\,\eqref{MTtime} to lower bound the time $T$ to implement the target gate if we can show that the evolution operator $U_{H_I}(T)$ for $H_I$ is capable of evolving some initial state to an orthogonal state. This is however complicated by the fact that $U_{H_1}(T)$ and $U_{H_2}(T)$ can be arbitrary due to the lack of constraints on $H_{1,2}(t)$.  Fortunately, we can use the fact that $U_{H_1}(T)$ and $U_{H_2}(T)$ are only arbitrary single-qudit unitaries to move forward. For concreteness, we use the iSWAP gate as our target gate $\mathcal{U}$ below as an example.

Let us choose an initial state of the form $|\psi_0\rangle = (c_0 |0\rangle + c_1 |1\rangle) \otimes |0\rangle$ and denote $|\psi_T\rangle = U_{H_I}(T)|\psi_0\rangle$. Using the identity $ U_{H_I}(T)= U_{H_1}^{\dagger}(T) U_{H_2}^{\dagger}(T) U_{H}(T)$, we find that
\begin{align}
\left|\langle\psi_0|\psi_T\rangle\right| &=\left|\langle\psi_0|U_{H_{1}}^{\dagger}(T)U_{H_{2}}^{\dagger}(T) U_{H}(T)|\psi_0\rangle\right| \nonumber \\
&\le\left|c_{0}\langle0|U_{H_{1}}^{\dagger}(T)|0\rangle+c_{1}\langle0|U_{H_{1}}^{\dagger}(T)|1\rangle\right|.
\end{align}
where we used that fact that $U_{H}(T)$ is an iSWAP gate in the two-qubit subspace. For any $U_{H_{1}}(T)$, we can then always choose appropriate values of $c_0$ and $c_1$ to make $\langle\psi_0|\psi_T\rangle=0$. In other words, there exists at least one initial state such that no matter how we choose $H_1(t)$, evolving under $H_I(t)$ for time $T$ results in an orthogonal state, provided that the evolution under $H(t)$ generates an iSWAP gate in the two-qubit subspace.

We then bound the energy uncertainty $\Delta E(t)$ for the Hamiltonian $H_I(t)$ at any time $t$ in any state $|\psi\rangle$ using
\begin{equation} \label{dEt}
   \Delta E(t) \le \sqrt{\langle H_I(t)^2 \rangle}  \le \sqrt{\lVert H_I(t)^2 \rVert} = \lVert H_I(t) \rVert \le J 
\end{equation}
where we use the fact that $\lVert H_I(t)^2 \rVert$ denotes the largest eigenvalue of $H_I(t)^2$, which cannot be smaller than $\langle H_I(t)^2 \rangle$. Combining Eq.\,\eqref{dEt} with Eq.\,\eqref{MTtime} above, we arrive at
\begin{equation} \label{theorybound}
   t_{\text{iSWAP}} \ge \frac{\pi}{2J}.
\end{equation}
Therefore, we obtain a theoretical speed limit for implementing a two-qubit iSWAP gate with a generic two-qudit Hamiltonian in Eq.\,\eqref{Ht} satisfying $\lVert H_c(t) \rVert \le J$. 

A similar but possibly more complicated analysis can be applied to obtain the speed limit for other two-qubit gates. For the iSWAP gate, we can further show that the speed limit given by Eq.\,\eqref{theorybound} is optimal. This can be seen first for the $d=2$ case, where we can generate the iSWAP gate directly using $H = J (\sigma_1^+ \sigma_2^- + \text{h.c.})$ that satisfies $\lVert H \rVert = J$. At time $T=\pi/(2J)$, we can easily check that $U_H(T)=e^{-i H T}=U_{\text{iSWAP}}$, exactly reaching the speed limit predicted by Eq.\,\eqref{theorybound}.

From Eq.\,\eqref{theorybound}, we also see that if $J \equiv \max \lVert H_c(t) \rVert $ is a constant independent of $d$, then no speedup can be obtained if we go from two qubits to two qudits. Fortunately, for many physical Hamiltonians, when we expand the computational space from qubits to qudits, the value of $J$ will increase naturally without physically increasing the interaction strength (such as moving the two qudits spatially closer). For example, the interaction between two capacitively coupled superconducting transmons (which are anharmonic oscillators) can be described by
\begin{equation} \label{Hctrans}
    H_c = g (a_1 + a_1^\dagger)(a_2 + a_2^\dagger)
\end{equation}
where $g$ characterizes the physical interaction strength and $a_i$ is the annihilation operator for the $i^{\text{th}}$ anharmonic oscillator. We can truncate the Hilbert space of each anharmonic oscillator to $d$ dimensions, making a system of two qudits. It is not hard to show that $\lVert H_c\rVert=gO(d)$. Intuitively, this is because the creation/annihilation operators have matrix elements of roughly $\sqrt{d}$ in the subspace formed by the two highest Fock states $|d-1\rangle$ and $|d-2\rangle$. Then Eq.\,\eqref{theorybound} predicts a potential $O(d)$ speedup for the implementation of a two-qubit gate, consistent with the numerical findings in Ref.\,\cite{Ashhab}.

Here we construct an explicit protocol based on the Hamiltonian in Eq.\,\eqref{Hctrans} to speedup a two-qubit iSWAP gate by a factor of $d-1$. We can replace the constant coupling strength $g$ in Eq.\,\eqref{Hctrans} by a parametric coupling $g \cos(\omega t)$ \cite{RayParam}, where $\omega$ is set to match the energy difference between the $|d-1,d-2\rangle$ and $|d-2,d-1\rangle$ energy eigenstates. If we assume the anharmonicity and frequency difference of the two transmons are sufficiently large compared to $g$, we induce an effective Hamiltonian of the form
\begin{equation} \label{Heff1}
    H_{\text{eff}} = g (d-1) ( \ket{d-1,d-2}\bra{d-2,d-1} + \text{h.c.}).
\end{equation}
If we further assume that single-qudit gates are arbitrarily fast, we can swap the information stored in the states $|d-2\rangle$ and $|d-1\rangle$ to the computational basis states $|0\rangle$ and $|1\rangle$ in no time, thus achieving an iSWAP gate in time $T=\pi/[2g(d-1)]$.

A very different way to obtain the same amount of speedup is to couple many states in the qudit space collectively to some states in the two-qubit subspace. For example, we can consider the following effective coupling Hamiltonian
\begin{equation}\label{Heff2}
    H_{\text{eff}}'= g (|0\rangle + |2\rangle + \cdots + |d-1\rangle)^{\otimes 2} \bra{11} + \text{h.c.}
\end{equation}
Simply speaking, we couple the state $|11\rangle$ to $(d-1)^2$ orthogonal Fock states with a uniform coupling strength $g$. If we start the two qudits in the state $|00\rangle$ and perform a single-qudit gate that takes $|0\rangle$ to $\frac{1}{\sqrt{d-1}}( |0\rangle + |2\rangle + \cdots + |d-1\rangle)$ for each qudit before and after the evolution of $ H_{\text{eff}}'$ for time $T= \pi/[2g(d-1)]$, we end up in the $|11\rangle$ state. A more careful analysis reveals that this process implements an iSWAP gate if we further flip the second qubit before and after the process.

Both protocols above saturate the theoretical speed limit given by Eq.\,\eqref{theorybound} in different ways and implement an iSWAP gate $O(d)$ faster than when the evolution is constrained to within the qubit subspace. Moreover, we can even combine the speedup obtained in the above two protocols in an experimentally realizable Hamiltonian for $d=3$. By parametrically driving the coupling $g$ in Eq.\,\eqref{Hctrans} with four different frequencies, corresponding to the energy differences between $\ket{00}$ and $\ket{11}$, $\ket{02}$ and $\ket{11}$, $\ket{20}$ and $\ket{11}$, as well as $\ket{22}$ and $\ket{11}$, we end up with the following effective coupling Hamiltonian in the rotating wave approximation:
\begin{equation}\label{Heff3}
     H_{\text{eff}}'' = g (\ket{00} + \sqrt{2} \ket{02} + \sqrt{2} \ket{20} + 2 \ket{22}) \bra{11} + \text{h.c.}
\end{equation}
It's not hard to show that an iSWAP gate can be achieved using the following gate sequence after projecting onto the two-qubit subspace:
\begin{equation}\label{Uprotocol}
    U_{\text{iSWAP}} = X_2 U_1^\dagger U_2^\dagger e^{-i H_{\text{eff}}'' \frac{\pi}{6g}} U_2 U_1X_2
\end{equation}
where $U_1$ and $U_2$ are rotations around the $y$ axis by an angle $2\arctan(\sqrt{2})$ for the effective qubit formed by the $\ket{0}$ and $\ket{2}$ states for qutrit 1 and qutrit 2 respectively, and $X_2$ flips the states $|0\rangle$ and $|1\rangle$ for the second qutrit. If we again assume arbitrarily fast single-qudit gates, we can now achieve an iSWAP gate in time $T= \pi/(6g)$, three times the speed for the same Hamiltonian restricted to the qubit subspace, where the speed limit is $\pi/(2g).$ We emphasize that this gate protocol also saturates the theoretical speed limit in Eq.\,\eqref{theorybound} due to $\lVert H_{\text{eff}}''\rVert = 3g$. Interestingly, we note that for $d=3$ the previous two protocols based on $H_{\text{eff}}$ in Eq.\,\eqref{Heff1} and $H_{\text{eff}}'$ in Eq.\,\eqref{Heff2} both give an iSWAP gate time $T = \pi/(4g)$ (assuming negligible single-qudit gate times). The protocol using $H_{\text{eff}}''$ is faster as it uses both stronger coupling strengths for higher excited states and a collective (but not uniform) coupling between states in the qudit subspace and a state in the qubit subspace.

\section{Optimal control results} \vspace{-5pt}

In practice the single-qudit gates, such as the $X_2, U_1, U_2$ in Eq.\,\eqref{Uprotocol}, cannot be arbitrarily fast. Driving the transitions in a qudit too strongly will off-resonantly drive other transitions in the qudit. It may also drive transitions outside the qudit subspace and transitions in the other qudit if individual addressing is not achieved by spatial separation, such as in most superconducting qubit platforms. Admittedly, these problems occur even if we are only driving the qubit transitions in usual gate designs, but the fact that we have to drive transitions outside the qubit subspace resonantly and strongly makes it much harder to control the errors caused by these off-resonant transitions. While we can control such errors using sufficiently weak drives, the long time we spend on single-qudit gates can easily negate any speedup we obtained for the gate protocols in Section II.

To overcome this challenge, we develop a quantum optimal control method based on the GRAPE algorithm \cite{GRAPE} to find optimal single-qudit Hamiltonians [$H_{1,2}(t)$ in Eq.\,\eqref{Ht}] that lead to a two-qubit gate in a coupled two-qudit system close to the theoretical limit studied in Section II with limited single-qudit drive strengths and the presence of off-resonant transitions. For simplicity, throughout the following discussion we assume the coupling Hamiltonian $H_c(t)$ in Eq.\,\eqref{Ht} is time-independent, but we can also optimize $H_c(t)$ if one can engineer the interaction Hamiltonian temporally.

The goal of our optimal control algorithm is to maximize the average fidelity $F$ between the evolution operator $U_H(T)=\mathcal{T}e^{-i\int^T_0 [H_1(t)+H_2(t)+H_c]dt}$ projected to the two-qubit subspace (denoted by $\mathcal{U}_H$ below) and the target two-qubit gate $\mathcal{U}$, defined by \cite{NielsenF}
\begin{equation}\label{F}
    F = \frac{1}{5} + \frac{1}{80} \sum_j \tr(\mathcal{U}U_j^\dagger \mathcal{U}^\dagger \mathcal{U}_H U_j \mathcal{U}_H^\dagger)
\end{equation}
where $U_j \in \{I,X,Y,Z\} \otimes \{I,X,Y,Z\}$ is one of the 16 two-qubit Pauli operators. To be specific, we focus on finding the fastest possible two-qubit iSWAP gate with a sufficiently high fidelity $F$ using two coupled qutrits with the interaction Hamiltonian given by Eq.\,\eqref{Heff3}. As we introduced in Section II, this Hamiltonian can be experimentally realized using parametrically coupled superconducting transmons \cite{RayParam}. To access the full space of two qutrits, we will drive resonantly the $\ket{0}-\ket{1}$ and the $\ket{1}-\ket{2}$ transitions for each qutrit and optimize the pulse envelopes of each drive as our control.

We first consider the practical scenario where single-qudit drives have finite strengths but off-resonant transitions can be ignored. This can be a good approximation if the anharmonicity and frequency difference of the two transmon qubits are much larger than the drive strengths and the interaction strength $g$. In this case, the drive Hamiltonians take the form
\begin{align}\label{drives}
    H_1(t) &= \left( \Omega_{1,1}(t)|0\rangle \langle 1| + \Omega_{1,2}(t)|1\rangle \langle 2| + \text{h.c.} \right) \otimes I  \nonumber \\
    H_2(t) &= I \otimes \left( \Omega_{2,1}(t)|0\rangle \langle 1| + \Omega_{2,2}(t)|1\rangle \langle 2| + \text{h.c.} \right)  
\end{align}
Here $\Omega_{i,k}(t)$ ($i,k=1,2$) represents the complex Rabi frequency of the drive that is resonant with the transition $\ket{k-1}-\ket{k}$ in the qutrit $i$. Note that here we do not include the diagonal Hamiltonian that defines the energies of the basis states $\ket{0},\ket{1},\ket{2}$ for each qutrit, as we are in the rotating frame of such a Hamiltonian.

\begin{figure}[h!]
    \hspace{-4pt}
    \includegraphics[width=0.45\textwidth]{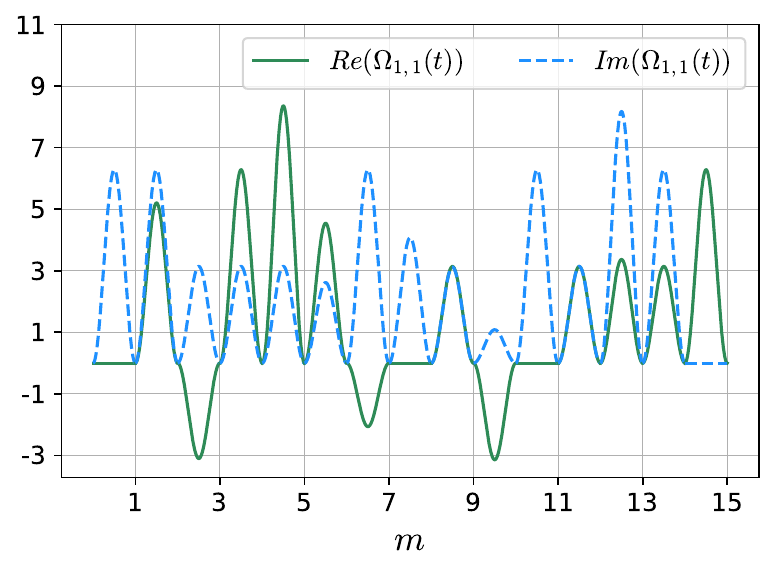}
    \caption{An example of the pulse shape $\Omega_{1,1}(t)$ (in units of $g$) in Eq.\,\eqref{Omegat} for the drive resonant with the $\ket{0}$-$\ket{1}$ transition of the first qudit. This pulse shape is obtained from our optimal control algorithm. For a better presentation, only the first 15 segments of the full 40-segment pulse are shown here.}
    \label{pulseEx}
\end{figure}

To perform numerical optimization efficiently, we need to discretize the continuous functions $\Omega_{i,k}(t)$ above. As in the standard GRAPE algorithm, we divide $\Omega_{i,k}(t)$ from $t=0$ to $t=T$ into $M$ segments each with a duration $\tau=T/M$. Within each segment, we assume $\Omega_{i,k}(t)$ is a slowly varying function that ensures $\Omega_{i,k}(t)$ is continuous across different segments. An example is:
\begin{equation}\label{Omegat}
    \Omega_{i,k}(t) = A_{i,k}^{(m)} \sin^2[\pi(m-1-t/\tau)], \, t\in [(m-1)\tau,m\tau]
\end{equation}
with $m=1,2,\cdots,M$. An illustration of $\Omega_{i,k}(t)$ in this form is shown in Fig.\,\ref{pulseEx}. We set $M=40$ in our calculations, which can be reached experimentally using commercial arbitrary waveform generators assuming a two-qubit gate time on the order of 10-100ns, typical for transmon qubits. Further increasing $M$ does not lead to any noticeable improvement in our results. We then optimize the $4M$ complex parameters given by $\{A_{i,k}^{(m)}\}$ via a gradient descent method with random initial values. We also run the optimization with multiple random initializations and choose the highest fidelity obtained from all runs.

\begin{figure}[h]
    \hspace{-10pt}
    \includegraphics[width=0.49\textwidth]{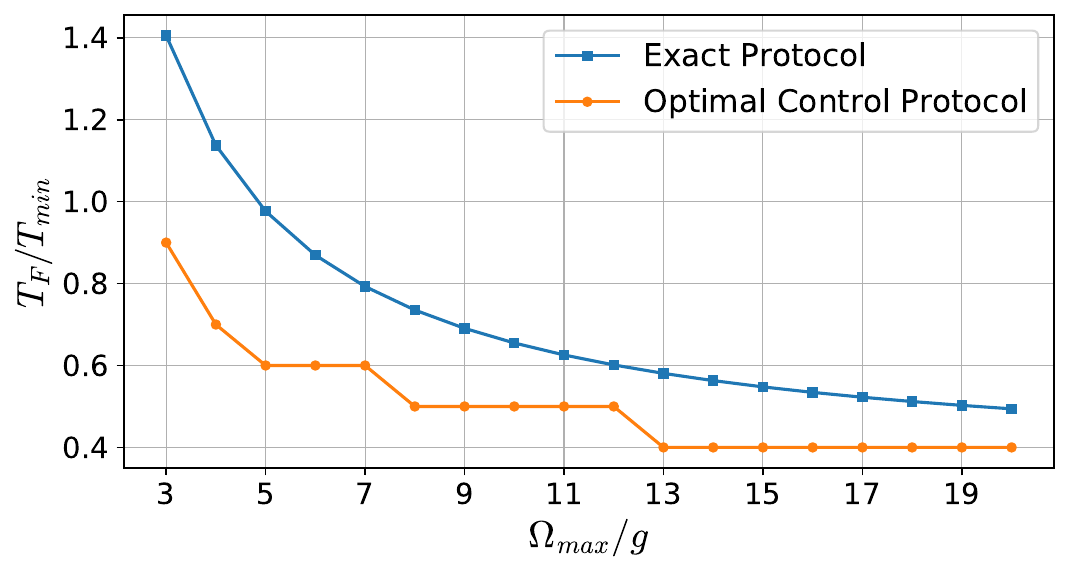}
    \caption{The minimum gate times $T_F$ for $F\ge 99.99\%$ achieved by the exact protocol in Eq.\,\eqref{Uprotocol} and the optimal control protocol without considering off-resonant transitions as a function of the maximum Rabi frequency $\Omega_{\text{max}}$ for the single-qudit drives.}
    \label{FvsOmax}
\end{figure}

To address the practical concern that single-qudit drives are of limited strengths, we apply a constraint $|A_{i,k}^{(m)}|\le \Omega_{\text{max}}$ for any $i,k,m$ in our optimization, implying the $\Omega_{\text{max}}$ is the maximum Rabi frequency of each drive. In Fig.\,\ref{FvsOmax}, we show the minimum gate time we obtained for a gate fidelity $F>99.99\%$ for different values of $\Omega_{\text{max}}$ (measured in units of the interaction strength $g$). We choose this fidelity threshold as it is higher than any experimental two-qubit gate achieved up to date. From Fig.\,\ref{FvsOmax}, one finds that if $\Omega_{\text{max}}/g\gtrsim 10$, a near-perfect iSWAP gate can be obtained for $T=0.4T_{\text{min}}$, which is close to the theoretical speed limit $T=T_{\text{min}}/3$. Here $T_{\text{min}} = \pi/(2g)$ is the minimum iSWAP gate time for the coupling Hamiltonian $H_c$ in Eq.\,\eqref{Heff3} restricted to the two-qubit subspace (i.e. $g(|01\rangle\langle 10|+\text{h.c.}$). Even for $\Omega_{\text{max}}$ as small as $3g$, we can still obtain a faster gate than the two-qubit case.

The speedup we obtained from our optimal control algorithm for a finite single-qudit drive strength is in fact non-trivial. If we instead follow the exact theoretical protocol in Eq.\,\eqref{Uprotocol}, we need to perform 4 single-qudit gates in serial and also spend a time $T_{\text{min}}/3$ for evolving $H_c$. Moreover, the single-qudit gates $U_{1,2}$ in Eq.\,\eqref{Uprotocol} require us to drive the transition $|0\rangle-|2\rangle$, which is a two-photon transition whose dipole moment is typically much smaller than that for the $\ket{0}-\ket{1}$ or the $\ket{1}-\ket{2}$ transitions in a transmon. Yet even if we assume that the maximum Rabi frequency of the $|0\rangle-|2\rangle$ drive is also $\Omega_{\text{max}}$, we still need to spend a minimum total gate time of
\begin{equation}
    T_{\text{exact}} =\frac{\pi + 2\arctan(\sqrt{2})}{\Omega_{\text{max}}} + \frac{\pi}{6g}
\end{equation}
to obtain an exact iSWAP gate in the two-qubit subspace following Eq.\,\eqref{Uprotocol}. As shown in Fig.\,\ref{FvsOmax}, we see that $T_{\text{exact}}$ is significantly larger than the gate time we obtained using the optimal control algorithm, despite that both gate times approach the theoretical limit $\pi/(6g)$ in the $\Omega_{\text{max}} \rightarrow \infty$ limit. The speed advantage brought by the optimal control method is likely due to the fact that single-qudit drives are applied simultaneously with the two-qubit interaction \cite{Joel}, while for the exact protocol the drives and the interaction are applied in serial. 

\begin{figure}[h]
    \hspace{-10pt}
    \includegraphics[width=0.48\textwidth]{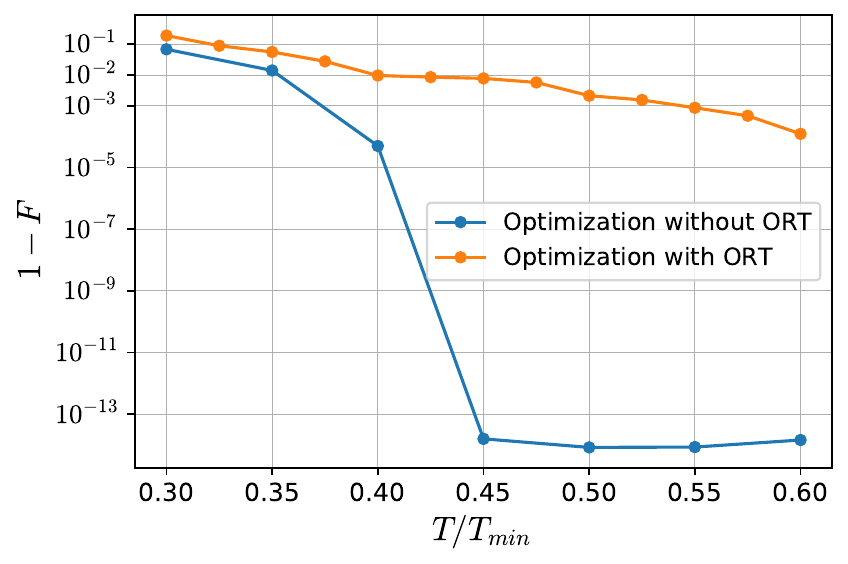}
    \caption{The infidelity $1-F$ of the iSWAP gate achieved by our optimal control algorithm with and without considering off-resonant transitions (ORT) as a function of the gate time $T$ using two coupled qutrits. The theoretical speed limit is given by $T=T_{\text{min}}/3$, which can only be achieved with infinitely fast single-qudit gates. Here we assume a maximum single-qudit Rabi frequency $\Omega_{\text{max}}=20g$ ($\Omega_{\text{max}}=40g$) for the optimization without (with) ORT.}
    \label{IFvsT}
\end{figure}

The optimal control algorithm can also give us much higher gate fidelity if we slow down the gate a bit. As shown in Fig.\,\ref{IFvsT}, we can achieve a gate infidelity below $10^{-13}$ for a gate time $T=0.45T_{\text{min}}$, which is still more than twice the speed of an iSWAP gate using only two qubits.

The results shown so far assume that the four drives in Eq.\,\eqref{drives} do not drive any other transitions off-resonantly. This assumption may not be valid if the two transmons are already strongly coupled, which is preferred for a fast two-qubit gate. For example, for a coupling strength $g$ in Eq.\,\eqref{Hctrans} on the order of $10$MHz, we need to drive each transmon with a maximum Rabi frequency on the order of $100$MHz in order to approach the theoretical speed limit as indicated by Fig.\,\ref{FvsOmax}. This is not much smaller than the anharmonicity of the transmon or the frequency difference between two transmon qubits (both typically a few hundred MHz \cite{Joel,RayParam}). Therefore, our next step is perform the optimization with the off-resonant transitions induced by the drives modelled by the Hamiltonians $H_{1,2}(t)$.

Since the qubit frequency of most transmons is much larger than their anharmonicity, we can still safely ignore any off-resonant two-photon transitions. For two qutrits, there are four single-photon transitions: the $\ket{0}-\ket{1}$ and $\ket{1}-\ket{2}$ transitions for each qutrit. Each of the four drives in Eq.\,\eqref{drives} is resonant with one of the four transitions, but can drive the remaining three transitions off-resonantly. In addition, we should also consider the fact that each of the four drives may also drive the transition $\ket{2}-\ket{3}$ off-resonantly for each transmon and excite the system outside the Hilbert space of two qutrits. To address this issue, we will expand our Hilbert space to $d=4$ to explicitly take into account the $\ket{3}$ state for each transmon. We then add a penalty proportional to the maximum and the average population outside the two-qutrit subspace to the loss function, i.e. the infidelity $1-F$. As we discuss in the Appendix, this penalty ensures that the leakage of the state outside the two-qutrit subspace causes a negligible (less than $0.1\%$) drop of the gate fidelity. Therefore, we can safely ignore off-resonant transitions involving even higher energy states, such as $\ket{4}$, $\ket{5}$, etc. 

Note that we can now set $\Omega_{\text{max}}/g$ to a larger value ($40$ from now on) as we have modelled all off-resonant transitions (5 for each drive tone). We set the anharmonicity to be $10g$ and the frequency difference between the two transmon qubits to be $15g$, which are  within the experimental reach for $g$ as large as 30MHz \cite{Joel,RayParam}. As shown in Fig.\,\ref{IFvsT}, our optimal control algorithm managed to find pulse shapes that lead to \textgreater $99\%$ gate fidelity at $T=0.4T_{\text{min}}$ and \textgreater $99.9\%$ gate fidelity at $T=0.55T_{\text{min}}$. While both the gate fidelity and gate speed are lower than the case without considering off-resonant transitions, we still obtain a notable speedup with a sufficiently high gate fidelity. For $g=30$MHz (in frequency not angular frequency), this means we can achieve an iSWAP gate in time $T\approx 4.6$ns with \textgreater $99.9\%$ fidelity, which is much faster than any two-qubit gate experimentally achieved with transmon qubits. In addition, we expect even higher fidelity and faster gate if the anharmonicity and the frequency difference between the two qubits are bigger.

Before we close this section, we point out a number of technical achievements we made in our optimal control algorithm. Our algorithm improves upon existing GRAPE based algorithms for quantum gate designs such as QuTiP \cite{qutip}, GRAPE-Tensorflow \cite{Schuster1}, and Juqbox \cite{LLNL1}. First, we use the state-of-art machine learning library Pytorch to perform auto-differentiation of the loss function over the parameters in the drive Hamiltonians. This auto-differentiation technique was also used in Ref.\,\cite{Schuster1}. Inspired by the state-of-art machine learning practice, we further employ a stochastic gradient descent algorithm with the Nesterov momentum method \cite{momentum} to improve the efficiency of the optimization process. In a recent work \cite{Joel}, we have shown that this machine learning inspired approach outperforms the standard GRAPE implementation (such as the one in the QuTiP software \cite{qutip} which uses the L-BFGS-B method) significantly when the parameter space is large and a high fidelity (e.g. \textgreater$99.9\%$) is needed. Second, we adopt a symplectic second-order Runge-Kutta (SRK2) method \cite{Feng2010} in finding the evolution operator $U_H(T)$ for the time-dependent Hamiltonian $H(t)$. A symplectic ordinary differential equation (ODE) solver conserves the unitarity of the calculated evolution operator, which allows us to use a larger step size to achieve a similar numerical accuracy, thus making the optimization more efficient. A similar symplectic method was used in Refs.\,\cite{LLNL1,LLNL2} also for GRAPE based algorithms.

We have made an open-source repository \cite{Basyildiz2023} that implements our optimal control algorithm in Python. This repository may benefit a wide range of researchers working on optimal quantum gate designs not only due to its high efficiency, but also due to its general framework that allows for multiple drives within or outside the two-qubit subspace and accounts for many off-resonant transitions. We have not seen such capabilities in any existing open-source repositories for optimal quantum gate designs \cite{qutip,Schuster1,Schuster2,LLNL1,LLNL2}.

\section{Conclusion and Outlook}\vspace{-5pt}
In this work, we have shown two primary results. First, we derived a theoretical speed limit for certain two-qubit gates implemented with two coupled qudits under the assumption of arbitrarily fast single-qudit gates. This speed limit is optimal and it can be exactly achieved with physical interactions among two qudits, providing an $O(d)$ speedup in the two-qubit gate time. In addition, we find that this speedup can come from naturally stronger coupling between higher energy states of the qudits, collective couplings between states in and outside the qubit subspace, or a combination of both. Also, we have constructed explicit protocols to demonstrate the speedup due to each of the above three cases.

Second, we addressed the concern that single-qudit gates cannot be arbitrarily fast in practice by developing a GRAPE based quantum optimal control algorithm and applying it to obtain an iSWAP gate in the two-qubit subspace close to the aforementioned theoretical speed limit for two coupled qutrits. We develop an open-source repository for our algorithm that implements state-of-art numerical optimization techniques, allows for multiple drive frequencies, and takes into account off-resonant transitions that cannot be ignored. We expect this repository to be widely useful for researchers working on optimal quantum gate designs, even if no states outside the qubit subspace is used intentionally.

An alternative way to speedup a two-qubit gate is to use one or more ancilla qubits, which is similar to the approach here in that both are using extra states to go beyond the two-qubit speed limit. However, it remains unclear where this approach can actually lead to a speedup. Finding out the speed limit of a two-qubit gate in a many-qubit system is also fundamental interesting as it relates to the study of Lieb-Robinson bounds and emergent locality in quantum many-body systems \cite{Chen2023}.

A natural follow-up to this work is to experimentally demonstrate the speedup we obtained for the iSWAP gate using two parametrically coupled transmons. In addition, we are also interested in designing specific protocols for other hardware platforms such as trapped ions to accelerate two-qubit gates there by using extra low-energy states that can be well controlled. Admittedly, these protocols will bring an extra level of complication in designing high-fidelity gates. But we believe such complication can be tackled with a combination of improvements in hardware and software, such as better individual addressing, more precise controls, longer coherence times, and more powerful optimal control methods. Our work thus creates a new pathway in further speeding up two-qubit gates when conventional approaches such as increasing the interaction strength between qubits have been exhausted.

\section*{Appendix: Leakage errors}

\begin{figure}[h]
    \hspace{-10pt}
    \includegraphics[width=0.45\textwidth]{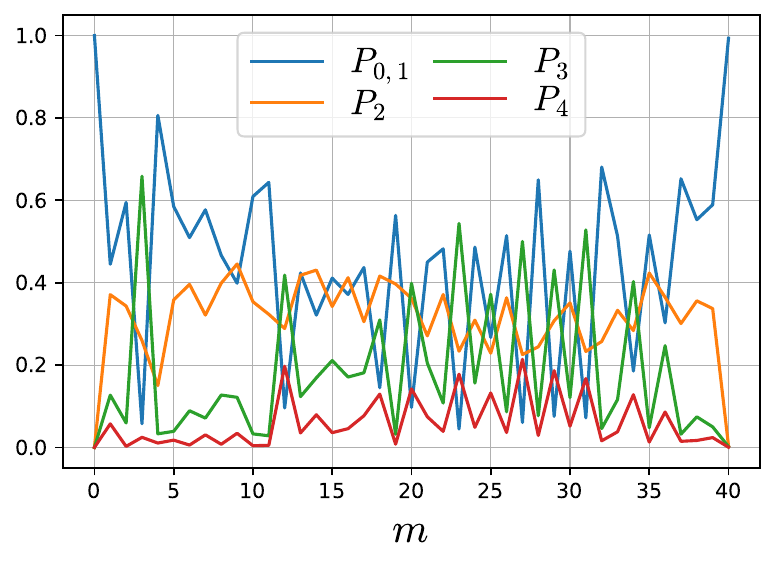}
    \caption{Probabilities $P_k$ for at least one transmon to be in the state $\ket{k}$ during the gate protocol obtained from the $T=0.5T_{\text{min}}$ point of the optimization with ORT curve in Fig.\,\ref{IFvsT}. $P_{0,1}$ denotes the probability in the two-qubit subspace.} 
    \label{Pexcited}
\end{figure}

In our optimal control algorithm discussed in Section II, we have modelled the two-qutrit system with four states ($\ket{0},\ket{1},\ket{2},\ket{3}$) to account for errors induced by possible leakage of the state outside the two-qutrit subspace for two transmons. We further add a penalty to our loss function in the optimization to suppress the population outside the two-qutrit subspace. To see how much leakage we still have, we calculate the probability $P_k$ ($k=2,3,\cdots$) for at least one transmon to be in the state $|k\rangle$ as well as the probability $P_{0,1}$ for both transmons to be in the two-qubit subspace. We start the system in one of the four computational basis states ($\ket{00}$, $\ket{01}$, $\ket{10}$, and $\ket{11}$) and plot in Fig.\,\ref{Pexcited} the probabilities $P_{0,1}$, $P_2$, $P_3$, and $P_4$ each averaged over the four initial states.

As we expected, over the application of $M=40$ pulse segments, $P_{0,1}$ starts at exactly 1 and ends at $0.9978$, showing that we have achieved a high-fidelity two-qubit gate in the subspace of two qudits with little leakage outside the qubit subspace. Due to the penalty we add to minimize excitations to states outside the two-qutrit subspace, we see that the population in the subspace involving state $|k\rangle$, denoted by $P_k$ decreases as we increase $k$. While the maximum value of $P_k$ is not negligible even for $k=4$, we note that the value of $P_k$ gets very close to zero at the end of the gate implementation.

\begin{figure}[h]
    \vspace{10pt}
    \hspace{-10pt}
    \includegraphics[width=0.48\textwidth]{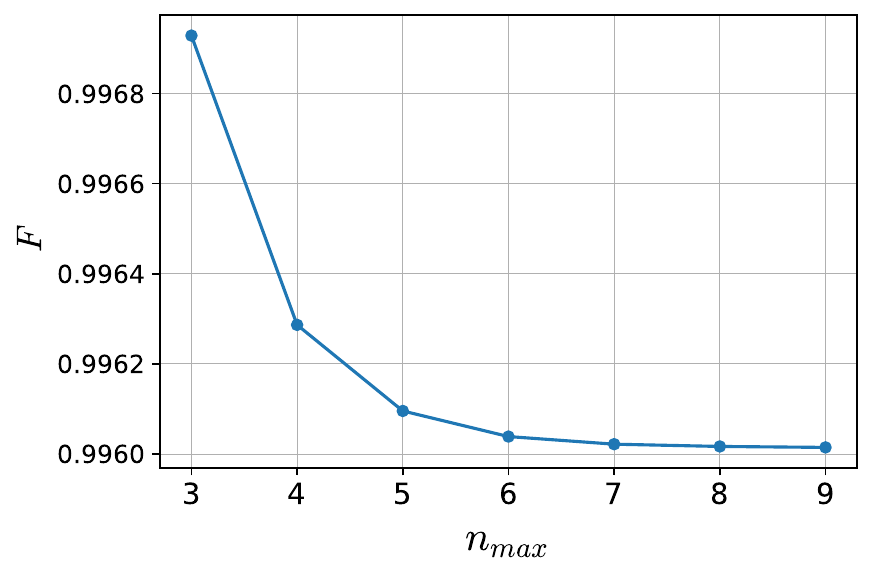}
    \caption{Gate fidelity $F$ to the ideal iSWAP gate when considering leakage to higher energy states due to off-resonant transitions. We consider all energy eigenstates of each transmon from $\ket{0}$ to $\ket{n_{\text{max}}}$. The particular gate calculated here is from the $T=0.5T_{\text{min}}$ point of the optimization with ORT curve in Fig.\,\ref{IFvsT}}
    \label{Fvsnmax}
\end{figure}

Fig.\,\ref{Fvsnmax} shows that the leakage to higher excited states during the gate implementation indeed has negligible effects on the gate fidelity obtained. As an example, we take the pulse shapes from the optimal control algorithm for a gate time $T=0.5T_{\text{min}}$ in the optimization with off-resonant transitions (see Section II and Fig.\,\ref{IFvsT}). We then expand our Hilbert space to consider all basis states from $\ket{0}$ to $\ket{n_{\text{max}}}$ for each transmon, and calculate the gate fidelity with off-resonant transitions to all higher excited states up to $\ket{n_{\text{max}}}$ taken into account. We see that the gate fidelity only decreases by less than $0.1\%$ as we increase $n_{\text{max}}$ from $3$ (used in the optimization algorithm) to $9$. In fact, the gate fidelities reported in Fig.\,\ref{IFvsT} for the optimization with ORT are already the fidelities calculated using $n_{\text{max}}=9$ (although the optimization is always done with $n_{\text{max}}=3$ for efficiency reasons). Thus we have properly taken into account the leakage errors.

\acknowledgements
We thank Dr. Raymond Simmonds, Tongyu Zhao, and Dr. Sahel Ashhab for enlightening discussions and the HPC center at the Colorado School of Mines for providing computational resources. We acknowledge funding support from the NSF NRT program under Grant. No. 2125899, the NSF RAISE program under Grant No. 1839232, and the W. M. Keck Foundation.

\bibliographystyle{apsrev4-1}
\bibliography{refs}

\end{document}